\documentclass[preprint]{agujournal2019}
\usepackage{url} 
\usepackage{lineno}
\usepackage{soul}
\usepackage{xcolor}

\draftfalse

\journalname{JGR: Space Physics}

\begin{document}

\title{Local acceleration of protons to 100 keV  in a quasi-parallel bow shock}

\authors{Krzysztof Stasiewicz\affil{1,2}, Bengt Eliasson\affil{3}, Ian J. Cohen\affil{4}, Drew L. Turner\affil{4}, and Robert E. Ergun\affil{5}}

\affiliation{1}{Space Research Centre, Polish Academy of Sciences, Warsaw, Poland}
\affiliation{2}{Department of Physics and Astronomy, University of Zielona G\'ora, Poland }
\affiliation{3}{SUPA, Department of Physics, University of Strathclyde, Glasgow, United Kingdom}
\affiliation{4}{The Johns Hopkins University,  Applied Physics Laboratory, Laurel, Maryland, USA}
\affiliation{5}{Laboratory for Atmospheric and Space Physics, University of Colorado, Boulder, USA}

\correspondingauthor{K. Stasiewicz}{krzy.stasiewicz@gmail.com}

 \begin{keypoints}
 \item	At quasi-parallel  shocks ions are rapidly accelerated from 10 eV to 100 keV by the ExB mechanism of electrostatic waves.
 \item The waves  are produced by  current-driven instabilities with frequencies between the proton and the electron cyclotron frequencies.
 \item The acceleration occurs locally within a few gyroperiods.
 \end{keypoints}

\begin{abstract}
Recent observations in the quasi-parallel bow shock  by the MMS spacecraft show rapid heating and acceleration of ions up to an energy of about 100 keV. It is demonstrated that a prominent acceleration mechanism is the nonlinear interaction with a spectrum of waves produced by gradient driven instabilities, including the lower hybrid drift (LHD) instability, modified two-stream (MTS) instability and electron cyclotron drift (ECD) instability. Test-particle simulations show that the observed spectrum of waves can rapidly accelerate protons up to a few hundreds keV  by the ExB mechanism. The ExB wave mechanism is  related to the surfatron mechanism at shocks but through the  coupling with the stochastic heating condition it produces significant acceleration on much shorter temporal and spatial scales  by the interaction with bursts of waves within a cyclotron period. The results of this paper are built on the heritage of four-point measurement techniques developed for the Cluster mission and imply that the concepts of Fermi acceleration, diffusive shock acceleration, and shock drift acceleration are not needed to explain proton acceleration to hundreds keV  at the Earth's bow shock.

\end{abstract}

\section*{Plain Language Summary}
The acceleration, or energization, of particles is a common and fundamental process throughout the universe. In particular, particle acceleration is a common occurrence at collisionless shocks that occur in plasmas in our solar system and beyond. This study presents new observations of the acceleration of protons to high energies (hundreds of keV) by waves at the bow shock upstream of the Earth, where the solar wind first encounters Earth's magnetic field. The observations by NASA's Magnetospheric Multiscale (MMS) mission  show a specific event where energized protons are measured, along with wave activity that is consistent with theoretical expectations for a certain type of interaction between the waves and particles that allows the waves to transfer energy to the protons. Numerical simulations are also performed and the results agree with what was observed in real life by the spacecraft. The results are important because they provide insight into potential processes that can create high-energy particles both near the Earth and at other astrophysical systems.

\section{Introduction} \label{seH1}
Collisionless shocks in the solar wind plasma are associated with particle heating and acceleration, and have over many years been the subjects of both spacecraft observations and theoretical investigations \cite{Ness:1964,Sagdeev:1966,Friedman:1971,Biskamp:1973,Bell:1978,Lee:1982,Wu:1984,Goodrich:1984,Blandford1987,Balikhin:1994,Gedalin:1995,Zank:1996,Krall:1997,Shapiro:2001,Eastwood:2005,Treumann:2009,Burgess:2012,Mozer:2013,Krasnoselskikh:2013,Guo2014,Wilson:2014,Park:2015,Kis:2018,Cohen:2019,Xu:2020}, to mention a few.
The physical mechanism for the plasma heating and acceleration at shocks has turned out to be elusive, and a consensus has not been reached despite 56 years of intense research since the discovery of the bow shock.

Recent observations by the Magnetospheric Multiscale Mission (MMS), which consists of four closely orbiting spacecraft \cite{Burch:2016} provided  high quality multipoint measurements in the bow shock suitable to address this problem. In a series of papers, \citeA{Stasiewicz:2020a,Stasiewicz:2020c,Stasiewicz:2020d,Stasiewicz:2021a} have shown that the heating and acceleration of ions and electrons at shocks of arbitrary configuration is related to current-driven instabilities that are initiated by the sharp plasma density gradients produced by the steepening of the shock waves. The diamagnetic currents caused by the density gradients initiate first the lower hybrid drift (LHD) instability \cite{Davidson:1977,Drake:1983,Gary:1993,Daughton:2003}, in the frequency range  $f_{cp}-f_{lh}$, between the proton cyclotron frequency $f_{cp}$, and the lower hybrid frequency $f_{lh}=(f_{cp}f_{ce})^{1/2}$, where $f_{ce}$ is the electron cyclotron frequency. The enhanced electric field of the LHD waves on short spatial scales produces fast ExB drift of electrons only, and the resulting cross-field current triggers the modified two-stream (MTS) instability \cite{Wu:1983,Umeda:2014,Muschietti:2017} in the frequency range $f_{lh}-f_{ce}$, and the electron cyclotron drift (ECD) instability having frequencies near multiples of the electron cyclotron frequency, $n f_{ce}$, where $n$ is an integer \cite{Lashmore:1973,Muschietti:2013,Janhunen:2018}.
The electric fields of these instabilities, $\sim 100\,$mV\,m$^{-1}$,  heat ions and electrons in a stochastic process, and can also accelerate  selected ions to hundreds keV  \cite{Stasiewicz:2021a}.

 The stochastic heating and acceleration of particle species with mass $m_j$ and charge $q_j$ ($j=e$ for electrons, $p$ for protons, $i$ for general ions) is controlled by the dimensionless function
\begin{equation}
\chi_j(t,\mathbf{r})  = \frac{ m_j}{q_j B^2} {\rm div}(\mathbf{E}_\perp); \quad |\chi_j|>1  \label{eq1}
\end{equation}
that depends on the mass-to-charge ratio and is also a measure of the local charge non-neutrality.  It is a generalization of the heating condition from earlier works \cite{Karney:1979,McChesney:1987,Balikhin:1993,Stasiewicz:2000,Vranjes:2010}, where the divergence was reduced to the directional gradient $\partial E_x/\partial x$.
 The particles are magnetized (adiabatic) for $|\chi_j|<1$,  demagnetized (subject to non-adiabatic heating) for $|\chi_j|>1$, and selectively accelerated to high perpendicular velocities when $|\chi_j|\gg1$ at frequencies $f\gg f_{cj}$. This mechanism has been already applied to heating of MeV ions observed at quasi-parallel shocks by Cluster \cite{Stasiewicz:2013}.

The analysis of MMS measurements supported by test-particle simulations presented in this paper demonstrates  that protons in the quasi-parallel bow shock  can be rapidly accelerated from 10 eV to a few hundred keV within a few gyroperiods by a mechanism controlled by Equation (\ref{eq1}).
The acceleration mechanism requires $\chi \gg 1$ and can bring some particles to  the gyration speed equal to the ExB drift velocity due to the wave electric field, i.e., to the speed $\widetilde{V}_E=\widetilde{E}_\perp/B$ \cite{Sugihara:1979,Dawson:1983,Ohsawa:1985}. The ExB mechanism is related to the surfatron mechanism at shocks \cite{Sagdeev:1966,Katsouleas:1983,Zank:1996,Ucer:2001,Shapiro:2003} but
works on much shorter temporal and spatial scales to reach significant energies by the interaction with bursts of waves \cite{Stasiewicz:2021a}.
Typical energies of ExB accelerated protons are a few hundred keV  in the quasi-parallel shock crossings investigated by the MMS  mission.

\section{Observations}

\begin{figure}[h]
\includegraphics[width=10cm]{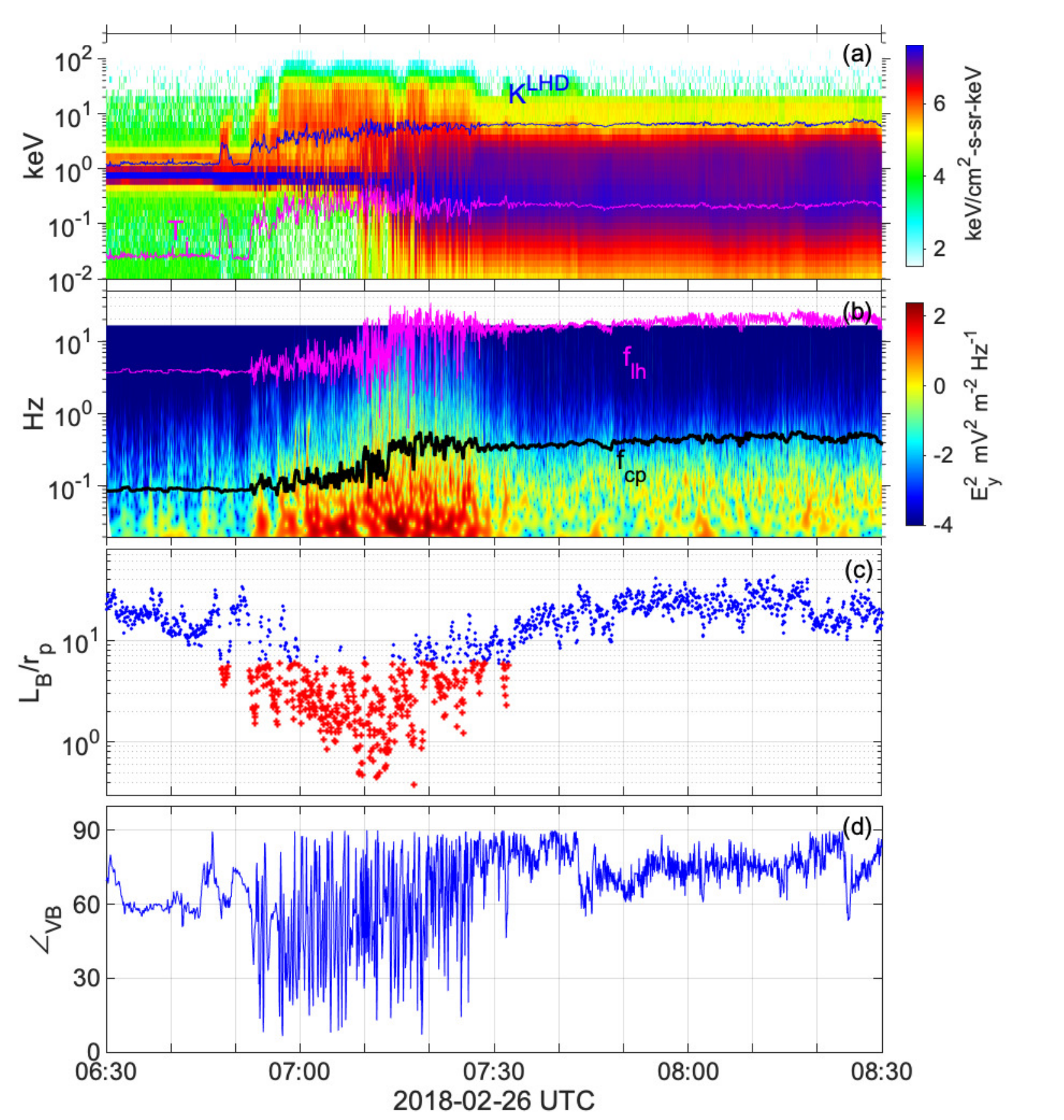}
\caption{MMS observations in the Solar wind region (06:30 - 06:50 UTC), followed by an extended quasi-parallel bow shock (06:55 - 07:25 UTC),  and the magnetosheath region (07:30 - 08:30 UTC). (a) Time versus energy spectrogram of the ion differential energy flux  measured by the MMS3 spacecraft. Overplotted is the acceleration capacity of the lower hybrid drift (LHD) waves given by Equation (\ref{LHD}) and the ion temperature $T_i$. The spectrogram combines Fast Plasma Investigation (FPI) data (10 eV - 20 keV) and Energetic Ion Spectrometer (EIS) data (20 - 300 keV).   (b) Time versus frequency spectrogram of the $E_y$ component of the electric field. Overplotted are the proton cyclotron frequency $f_{cp}$ and the lower hybrid frequency $f_{lh}$. The power spectrum covers frequencies 0.02 - 16 Hz.  (c)  The  gradient scale of the magnetic field $L_B$ normalized by the thermal proton gyroradius $r_p$. Red color marks regions LHD unstable. (d) Angle between the magnetic field  and the ion flow velocity. Active heating and acceleration of ions to 100 keV occurs during quasi-parallel configuration (06:55 - 07:25 UTC).  \label{Fig1}}
\end{figure}

On February 26, 2018 the  four MMS spacecraft  crossed  the bow shock and entered the  magnetosheath on an inbound orbit, where they made observations shown in Figure \ref{Fig1}.  The time versus energy spectrogram of the ion differential energy flux  (panel (a)) is produced with measurements from  the Fast Plasma Investigation (FPI) \cite{Pollock:2016} in the energy range 10 eV - 20 keV, and from the Energetic Ion Spectrometer (EIS) \cite{Mauk:2016} in the energy range 20 - 300 keV.  The spacecraft was in the solar wind at 06:30 UTC at position (13.1,  -4.4,  4.4) $R_E$ geocentric solar ecliptic (GSE) coordinates, crossed an extended quasi-parallel bow shock  during times  06:55 - 07:25 UTC,  and continued path through the magnetosheath until 08:30 UTC at position (10.9,  -2.5,  3.5) $R_E$ GSE.  The value of the Alfv\'en Mach number was $M_A\approx 7$ at 06:30, dropped to $M_A\approx 1.5$ at 07:20, and to $M_A\approx 0.9$  after 07:30 in the magnetosheath. Overplotted in panel (a) is the ion temperature obtained from the moments of the distribution function measured by FPI, and the acceleration capacity of the LHD waves for protons,  expressed as \cite{Stasiewicz:2021a}
\begin{equation}
K^{\mathrm{LHD}} \approx 1.5 \left(\frac{m_p}{m_e} T_eT_p\right)^{1/2}, \label{LHD}
\end{equation}
where $T_e,\,T_p$ are electron and proton temperatures.
The expression (\ref{LHD}) constitutes the limiting value for proton acceleration by lower hybrid drift waves generated on density gradients observed at shocks.  Protons can be accelerated resonantly by LHD waves to energies $K\sim m_p v_{ph}^2$ where $v_{ph}$ is the phase velocity of the LHD wave. This limit is generally  observed by the MMS spacecraft when it crosses  quasi-perpendicular shocks ($K<10\,$keV), while in quasi-parallel shocks protons are accelerated locally to hundreds keV, which is seen during time interval  06:55-07:25 UTC. These general limits can be easily verified by checking the Quicklook plots for the MMS mission (see the Web address provided in the Acknowledgements.)

Panel (b) in Figure \ref{Fig1} shows the time versus frequency spectrogram of the $E_y$ GSE component of the electric field measured by the electric field experiment \cite{Ergun:2016,Lindqvist:2016} on MMS3.  The spectrogram is produced with continuous wavelet transform using the software available at the Web address provided in the Acknowledgements.
Overplotted are the proton cyclotron frequency $f_{cp}$ and the lower hybrid frequency $f_{lh}$. The power spectrum covers frequencies 0.02 - 16 Hz. The magnetic profile of the shock can be seen in the plot of $f_{lh}\propto B$, computed from the full time resolution measurements of the magnetic field \cite{Russell:2016}. Waves below $f_{cp}$ are classified as magnetosonic waves, while waves between $f_{cp}-f_{lh}$ can be associated with LHD waves generated by the density gradients. Other wave modes like whistlers and magnetosonic waves are also likely to occur in this frequency range.

Panel (c)  shows the  scale of the magnetic field gradient $L_B= B|\nabla B|^{-1}$  derived from four point measurements  using the method of  \citeA{Harvey:1998} developed for the Cluster mission. It is normalized with the thermal proton gyroradius $r_p=(2T_i/m_p)^{1/2}/\omega_{cp}$ with $\omega_{cp}=eB/m_p$ being the proton angular cyclotron frequency. The scale $L_B$ is used here as a proxy for the scale $L_N= N|\nabla N|^{-1}$ of the density gradient, because the derivation of the $L_B$ scale is more reliable than $L_N$ in the solar wind. In shock compressions and in the magnetosheath these two  scales coincide due to the plasma being frozen in to the magnetic field,  while there are some differences in the solar wind
 \cite{Stasiewicz:2020a,Stasiewicz:2020c,Stasiewicz:2020d,Stasiewicz:2021a}. 
Red color in panel (c) marks regions where $L_B/r_p \approx L_N/r_p< (m_p/m_e)^{1/4}\approx 6$, which corresponds to the LHD instability  condition derived some 40 years ago \cite{Davidson:1977,Drake:1983}. It is remarkable that this limit is observed in MMS measurements, where the LHD waves in panel (b) and high energy protons in panel (a) coincide with the unstable, red regions in panel (c).

Panel (d) shows the angle between the magnetic field  vector and the ion bulk flow velocity $\mathbf{V}_i$, denoted as $\angle_{VB}$. The antiparallel values are here converted to 0-90 degrees range.
It is generally thought that the angle between the shock normal  direction and the interplanetary magnetic field  $\angle_{NB}$ controls the character of the shock, and the energization and heating processes of the plasma.  The shock normal direction is usually determined from the Rankine-Hugoniot equations \cite{Vinas:1986} or the minimum variance analysis of the magnetic field under assumption of the planar geometry \cite{Sonnerup:1998}, but results of such analyses are scale and time dependent, and give unreliable values for turbulent shocks.
An easy alternative is based on  the angle between the  geocentric radius vector to the satellite and the magnetic field, $\angle_{RB}$,  which may give reasonable values in  the dayside bow shock within some cone around the sunward direction. However, we find that the angle  $\angle_{VB}$ is  a better indicator of the character of the shock.  It is a physical (not geometrical) parameter and can be easily determined in any position on the satellite trajectory. This angle indicates a quasi-parallel character of the shock during times  06:55-07:25 UTC with quasi-parallel plasma flows $\angle_{VB}< 45^\circ$ seen in panel (d).
The local acceleration of protons to 100 keV  observed in panel (a) coincides with LHD waves in panel (b) and with strong magnetic field and the density gradients,  as indicated by  scales $L_N\sim L_B< 6r_p$ in panel (c).

\begin{figure}[h]
\includegraphics[width=10cm]{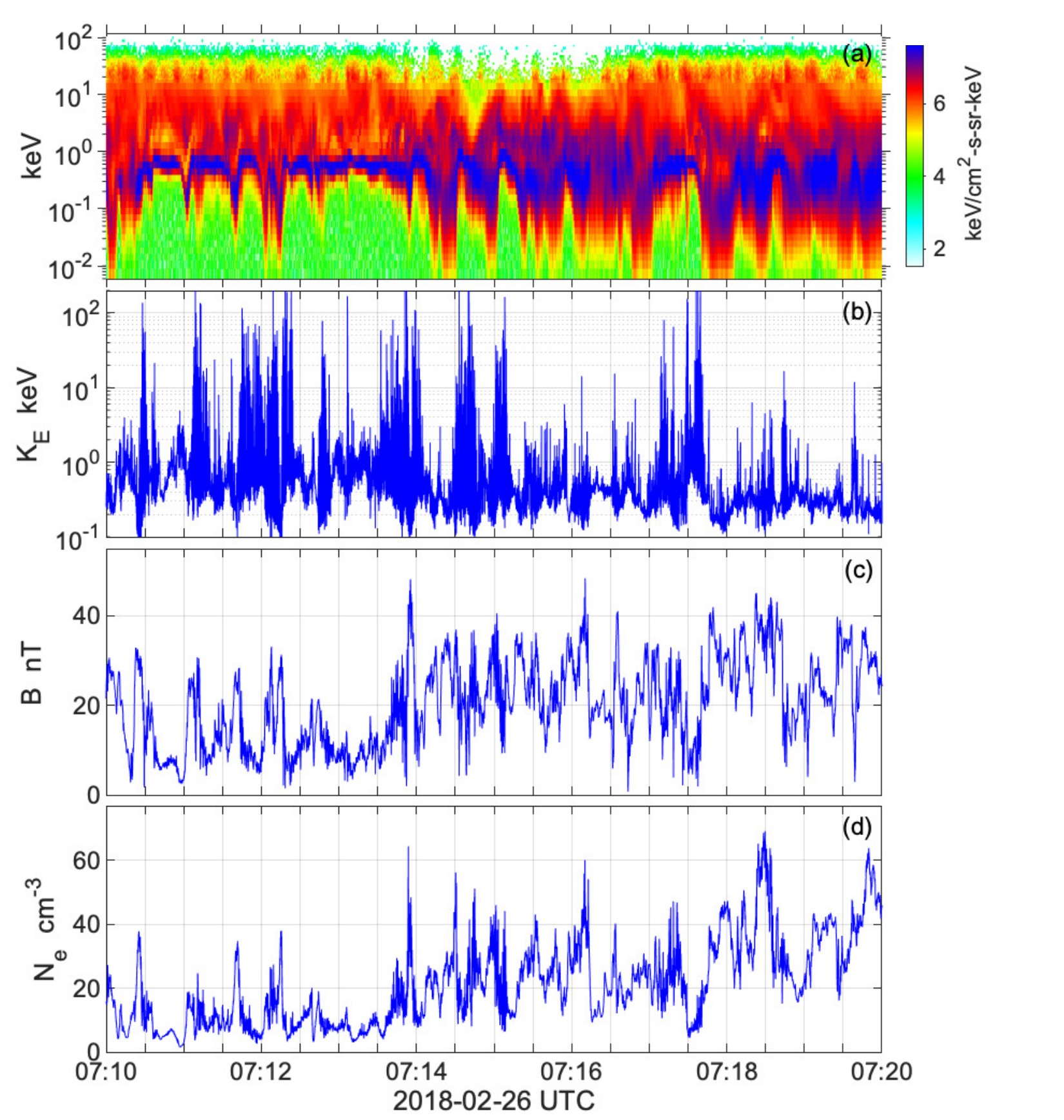}
\caption{High-time resolution burst data during 10 min of Figure \ref{Fig1} that coincides with the quasi-parallel shock region. (a) Time versus energy spectrogram of the ion differential energy flux from  the Fast Plasma Investigation (FPI) and the Energetic Ion Spectrometer (EIS) measured by the MMS3 spacecraft. (b)  The energization capacity $K_E$ of waves in the frequency range 0-512 Hz given by Equation (\ref{KE}). (c) The modulus of the magnetic field $B$.    (d) The electron number density measured by FPI.  \label{Fig2}}
\end{figure}

While Figure~\ref{Fig1} shows data from the lower time-resolution survey mode, Figure~\ref{Fig2} shows 10 minutes of high-resolution burst-mode data from the quasi-parallel shock region of Figure~\ref{Fig1}. In panels (c) and (d) in Figure \ref{Fig2} we see a series of shocklets, i.e. compressions of the magnetic field and plasma density, typical for quasi-parallel shocks. These compressions are related to strong gradients of the magnetic field, with the scale length $L_N\approx L_B<6r_p$ shown in Figure~\ref{Fig1}(c). As mentioned above and discussed in great detail in other papers \cite{Stasiewicz:2020c,Stasiewicz:2020d,Stasiewicz:2021a} these gradients destabilize first the LHD instability \cite{Davidson:1977,Drake:1983,Gary:1993,Daughton:2003}, which creates wave structures with enhanced electric fields and perpendicular scales between the ion and electron gyroradii. The ExB drift of electrons only in these small-scale structures creates cross-field currents that trigger the MTS instability \cite{Wu:1983,Umeda:2014,Muschietti:2017}, and the ECD instability \cite{Lashmore:1973,Muschietti:2013,Janhunen:2018}.
The electric field of these instabilities extend above the limit 16 Hz obtained from survey mode data shown in Figure~\ref{Fig1}(b). Higher frequency wave modes can be investigated with burst data of the electric field  sampled at the rate 8192 s$^{-1}$.  Field amplitudes of waves in the ECD frequency range $f> 500$ Hz reach $\sim 200$ mV\,m$^{-1}$.
During this time interval the mean value of the proton thermal gyroradius was $r_{p}\approx160\,$km, electron gyroradius $r_e\approx 1.2\,$km. The ion plasma beta was $\beta_i\approx 8$, and $\beta_e\approx 0.8$.

It has been shown by \citeA{Sugihara:1979,Dawson:1983,Stasiewicz:2021a} that  particles in large amplitude electrostatic waves can be accelerated  to the  ExB velocity computed with the wave electric field, i.e., $\widetilde{V}_E=\widetilde{E}_\perp /B$. The energization capacity for protons is found to be \cite{Stasiewicz:2021a}
\begin{equation}
K_{E}^p\approx \frac{m_p}{2} [v_{\perp 0}^2 + (\widetilde{E}_\perp/B)^2],    \label{KE}
\end{equation}
where $v_{\perp 0}$ is the initial thermal speed $v_{Tp}$, or the resonant velocity  $v_{ph}$ of protons subject to acceleration by waves.
This acceleration mechanism is related to $\chi$ in Equation (\ref{eq1}) and accelerates particles to the ExB velocity, which has led to its name
 '$\chi$-acceleration', or 'ExB wave acceleration' \cite{Stasiewicz:2021a}.
The value of $K_E^p$ computed  for frequencies $f<512$ Hz and for the ion thermal speed $v_{\perp 0}=v_{Ti}$ is shown in Figure~\ref{Fig2}(b). It shows the acceleration capacity of waves at the level of 200 keV, which corresponds well to the  maximum energy of protons in panel (a).   This indicates that these waves could indeed  be the source of energy for the  measured protons.

\begin{figure}[t]
\includegraphics[width=10cm]{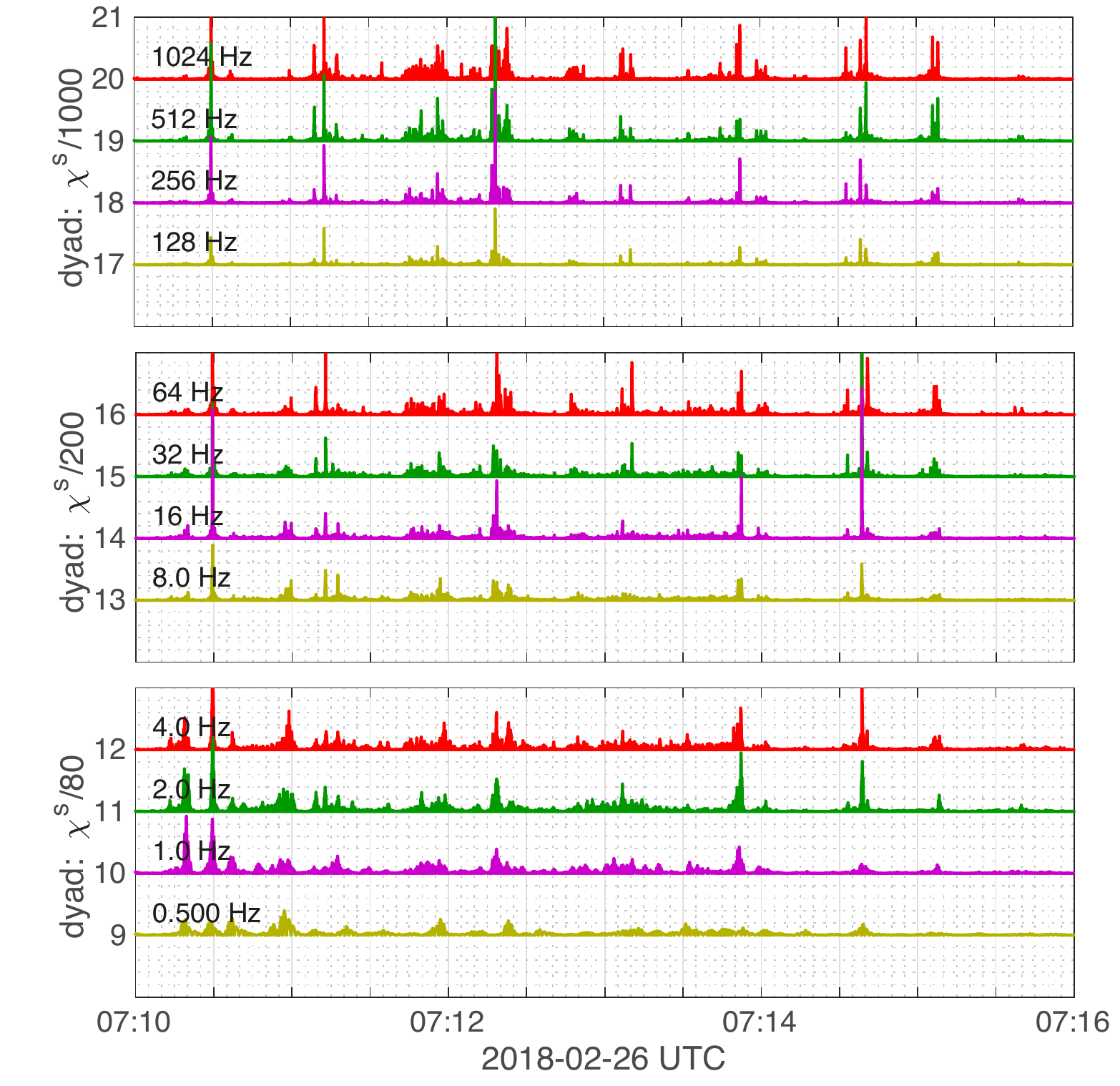}
\caption{Decomposition of $\chi_p$ given by Equation (\ref{eq1}) into discrete frequency dyads $\chi^s$  for a interval of Figure~\ref{Fig2}. \label{Fig3}}
\end{figure}

\begin{figure}[h]
\includegraphics[width=12cm]{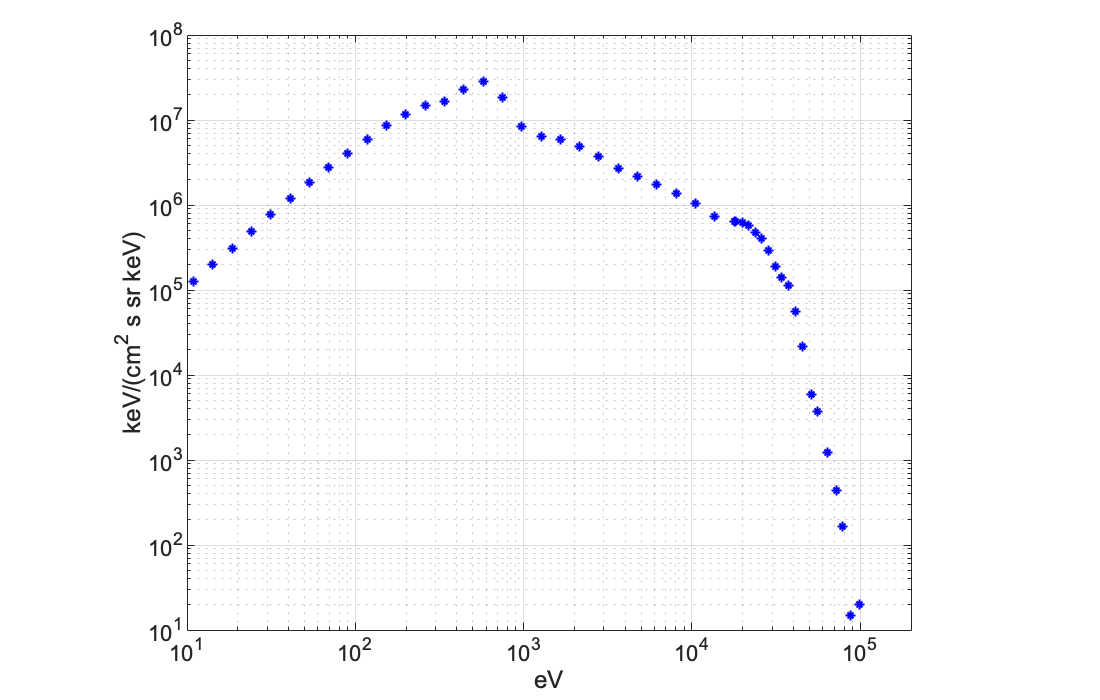}
\caption{The averaged omnidirectional ion differential  energy flux distribution measured by the Fast Plasma Investigation (FPI) and the Energetic Ion Spectrometer (EIS) instruments during the time interval 07:10--07:20  in Fig.~\ref{Fig2}(a).  \label{Fig4}}
\end{figure}

To check whether the measured waves are capable to stochastically accelerate protons we compute expression (\ref{eq1}) from 4-point measurements \cite{Harvey:1998}.
The computed $\chi_p(t,r)$ is  decomposed into discrete frequency bands  with orthogonal wavelets \cite{Mallat:1999}.   This technique is very different from band-pass filtering. The signal is divided into discrete frequency layers (dyads) that form $f=2^{-s}f_N$ hierarchy  ($s=0,1,2,...$) starting from the Nyquist frequency ($f_N$ is half of the sampling frequency). The decomposition is exact, i.e., the sum of all components gives the original signal, and the orthogonality means that the time integral of product of any pair dyads of different frequencies is zero.
The amplitudes of the decomposed signal are shown in Figure~\ref{Fig3} in the frequency range 0.5 -- 1024 Hz. We see  values varying from  $\chi^s\sim 20$  for $f=0.5\,$Hz to  $\chi^s\sim 2000$  for $f=1024\,$Hz, well above the stochastic threshold of $\chi^s\sim 1$. In other words, in the measured wave fields, protons are strongly demagnetized and subject to stochastic heating. The proton cyclotron frequency in this time interval was $f_{cp}=0.30\,\pm 0.15$ (standard deviation) Hz. The lower hybrid frequency was  $f_{lh}=13\,\pm 6\,$Hz, and the electron cyclotron  $f_{ce}=550\,\pm 250\,$Hz.  Figure \ref{Fig3} is also a good illustration of the cascade of instabilities that start from LHD at the bottom frequencies and progress through MTS and ECD to higher frequencies.
The measured waves with peak amplitudes of 200  mV\,m$^{-1}$ are likely responsible for producing the ion differential energy flux distribution shown in Figure~\ref{Fig4}, which is the time-average of the distribution in Figure~\ref{Fig2}(a). Characteristic features of Figure \ref{Fig4} is a power law distribution for low energy ions up to about 300 eV, the peak near 0.7 keV representing the flow energy of the solar wind ions, followed by a power law spectrum up to about 30 keV, and a sharp drop with a few ions reaching  100 keV. We assume here that the ions are protons.

 The ExB mechanism has some components in common with the surfatron mechanism introduced by \citeA{Katsouleas:1983} for the relativistic acceleration of electrons in laser plasmas.
The surfing acceleration, as explained by \citeA{Shapiro:2001,Shapiro:2003}, applies to quasi-perpendicular shocks, where electrostatic waves propagate in the sunward, $x$-direction, while the particles are accelerated in the $y$-direction, tangentially to the shock front. The surfing mechanism requires wide front of coherent waves and acceleration is done after multiple ion reflections between the shock and the upstream region \cite{Zank:1996,Shapiro:2001}. The acceleration is provided mainly by the macroscopic convection electric field $E_y$ when particles surf along the shock front, and partly by wave fields for trapped particles.
In contradistinction to the cited models of surfing acceleration, the  ExB wave mechanism does not require extended surfing because it is coupled with the stochastic condition  (\ref{eq1}). It works on turbulent wave fields  with wide frequency spectrum as shown further in this paper.
It leads to ion acceleration not only at shocks, but in all types of magnetized plasma environments, provided there are waves or structures that satisfy $|\chi |\gg1$.  The ExB acceleration is done exclusively by the wave electric field during times shorter than the cyclotron period. The wave electric fields in shocks  at frequencies $f\gg f_{lh}$ are typically 20 times larger than the dc convection fields, which ensures efficient acceleration by the ExB wave mechanism.

Waves below the lower hybrid frequency are associated with large amplitude magnetic fluctuations which is observed also in the vicinity of the reconnection regions \cite{Vaivads:2004,Norgren:2012,Ergun:2019,Ergun:2019a,Graham:2019}.  The electromagnetic component could be related to   ion  whistler waves created in the  density striations by mode conversion from LHD waves \cite{Rosenberg:2001,Eliasson:2008,Camporeale:2012}. Recently,  \citeA{Ergun:2019,Ergun:2019a} have suggested that these are electromagnetic drift waves, possibly related to electron vorticies. The exact nature of these fluctuations does not affect the conclusions of this paper. Simulations made by  \citeA{Lembege:1983,Lembege:1984} and \citeA{Ohsawa:1985} indicate that electromagnetic  waves in the frequency range $f_{cp}-f_{lh}$ are equally efficient ion accelerators as the electrostatic waves.

In the next section we show that the stochastic mechanism can indeed bring the solar wind protons with  temperature of 10 eV  to the observed energy of 100 keV at the shock crossing as seen in Figure~\ref{Fig1}(a).

\section{Test-particle simulations}

To demonstrate that the observed waves are capable to create the proton differential energy flux distribution in Figure~\ref{Fig4} we use an extension of the simulation model described in earlier papers \cite{Stasiewicz:2020c,Stasiewicz:2020d,Stasiewicz:2021a} to include a wide spectrum of waves.
  In the  magnetic field $\mathbf{B}_0=(0,0,B_0)$ there is  a macroscopic convection electric field $E_{y0}$ that drives particles into an  electrostatic wave turbulence composed of a number $s=1,2,3,...$ of harmonic waves
 $E_x^s=E_{x0}^s\sin( \omega_{D}^s t -k_x^s x  +\varphi^s)$ with wavenumbers $k_x^s=2\pi/\lambda^s$, random phases $\varphi^s$, and the Doppler shifted frequencies $\omega_{D}^s$  in the spacecraft frame. These wave modes could correspond to waves that produce the spectrum of $\chi_p$ in Figure~\ref{Fig3}.

Trajectories and velocities of particles  with mass $m$, charge $q$  are  determined by the Lorentz equation $md\mathbf{v}/dt=q(\mathbf{E}+ \mathbf{v}\times \mathbf{B}_0)$.
We use dimensionless variables with time normalized by $\omega_c^{-1}$, space by $(k_x^0)^{-1}$ and velocity by $\omega_c/k_x^0$ with $\omega_c=q B_0/m$ being the angular cyclotron frequency.  Superscript $s=0$ in $k_x^0$ denotes the reference wave mode in the simulations that is used for normalization of space dimension and velocities.
Changing to the plasma rest frame, the following system of equations is obtained
\begin{eqnarray}
   \frac{du_x}{dt}&= &\sum_s \frac{k_x^0}{k_x^s} \chi^s \sin\bigg[  \Omega^s t -  \frac{k_x^s}{k_x^0} x +\varphi^s\bigg] + u_y,\label{eqh1} \\
   \frac{du_y}{dt}&=& -u_x,\label{eq3}\\
  \frac{d x}{dt}&=&u_x,\\
  \frac{d y}{dt}&=&u_y,\label{eqh2}
\end{eqnarray}
where the normalized wave frequencies in the plasma frame  are
$\Omega^s =(\omega_D^s -k_x^s E_{y0}/B_0)/ \omega_c $,
and
\begin{equation}
 \chi^s  =\frac{E_{x0}^s}{B_0}\frac{k_x^s}{\omega_c}
 \end{equation}
corresponds to the stochastic heating parameter (\ref{eq1}) for a single wave mode.  This is in fact the normalized amplitude of the wave induced ExB drift speed $V_{E,y}^s=-E_x^s/B_0$, not to be confused with the convection drift $V_{E,x}=E_{y0}/B_0$ that is absorbed in the normalized frequency $\Omega$. The parallel acceleration along the magnetic field lines is not considered here; however, it has been found that the parallel component of the wave electric field effectively leads to the isotropisation of the distribution of particles initially accelerated in the perpendicular direction \cite{Stasiewicz:2020d}.

A test-particle simulation using $M=10^7$ particles to resolve the proton velocity distribution function is carried out for equations (\ref{eqh1})-(\ref{eqh2}) in the plasma reference frame. The initial distribution function is a 3D Maxwellian  $F(x,y,v_x,v_y,v_z)$ with number density $n_p=10\,{\rm cm^{-3}}$  that corresponds  to the data in the beginning of Figure~\ref{Fig2}, and proton temperature 10 eV. The initial gyration velocity of  a particle  is $(v_{x0},v_{y0},0)$, and  $v_{\perp 0}=(v_{x0}^2+v_{y0}^2)^{1/2}$. In normalized variables  it becomes
\begin{equation}
u_{\perp 0}=k_\perp v_{\perp 0}/\omega_c=k_\perp r_{c} \label{u0}
\end{equation}
with the initial Larmor radius $r_{c}=v_{\perp {0}}/\omega_c$, and $k_\perp=k_x^0$.

To link the dimensionless simulation variables, and in particular the initial value for  the thermal proton gyroradius $k_\perp r_c$, where $r_c=v_{Tp}/\omega_{cp}$ and $v_{Tp}=(2T_p/m_p)^{1/2}$, with physical parameters we apply the following reasoning. According to the results of \citeA{Daughton:2003} and \citeA{Umeda:2014} the dominant LHD wave mode has $k_\perp (r_e r_i)^{1/2}\sim1$. This assumption has also been used to derive equation (\ref{LHD}), which appears to hold in shocks \cite{Stasiewicz:2021a}. Because $r_e\approx r_i/100$ in case of Figure~\ref{Fig2} we obtain $k_\perp r_i\sim 10$ for the LHD waves. The protons have a temperature $\sim 250\,$eV in Figure~\ref{Fig2}, which is  25 times higher than the cold solar wind protons.  This leads to  the initial value of $k_\perp r_c=2$ for  solar wind protons at  temperature $T_i=10$ eV as the starting point for the simulation.  Because the initial  $k_\perp r_c =u_{T\perp}$  equals the dimensionless initial thermal speed by virtue of Equation (\ref{u0}), this condition is also used to translate dimensionless velocities $u_{x,y}$  to physical values $v_{x,y}$ and to  the energies of accelerated particles.
\begin{figure}[h]
\includegraphics[width=9.5cm]{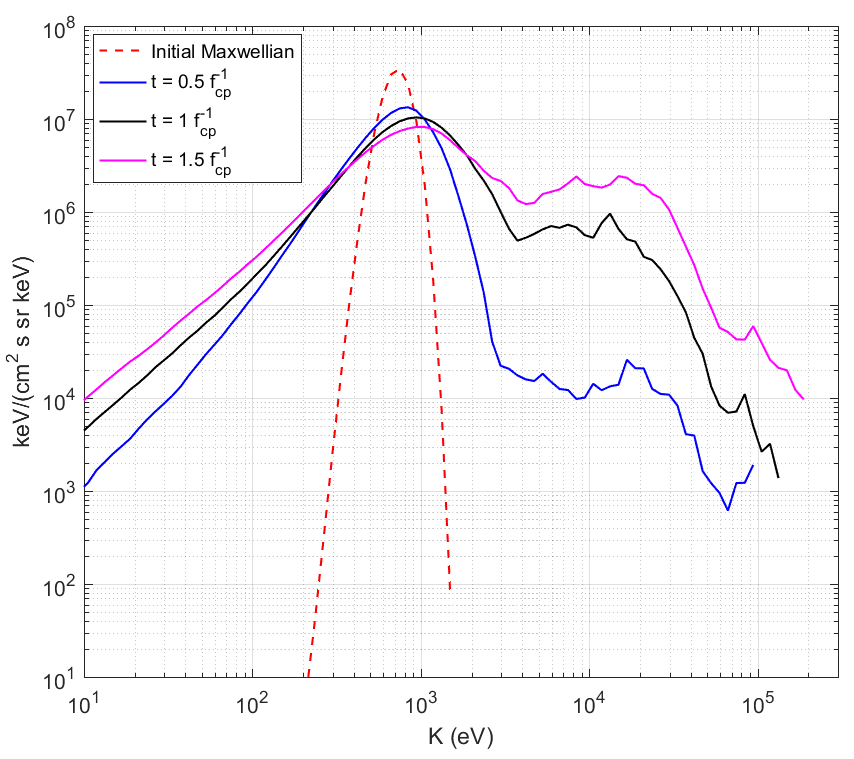}
\caption{Simulated differential proton energy flux distribution. The initial distribution function is a Maxwellian with number density $n_p$=10\,cm$^{-3}$ and proton temperature 10 eV shifted in velocity  by $v_{x0}$= 360\,km\,s$^{-1}$, which corresponds to the proton streaming energy of 0.7 keV. The magnetic field is taken to be 20 nT, corresponding to $f_{cp}$=0.3\,Hz. There are 9 interacting waves with frequencies: $\Omega^s$=[2, 4, 8, 16, 32, 64, 128, 256, 512 Hz]$f_{cp}^{-1}$,  amplitudes $\chi^s$=[5, 10, 30, 120, 400, 1000, 1600, 2000, 3000] ,  and wavevectors $k_x^s/k_x^0$=[0.2, 0.5, 0.8, 1, 2, 3, 4, 5, 10], where $k_x^0$ corresponds to the wave $s$=4 at 16 Hz. Different curves correspond to simulation times: 0, 0.5, 1, and 1.5 gyroperiods.  }
\label{Fig5}
\end{figure}

We run simulations using 9 waves with frequencies and amplitudes taken from Figure~\ref{Fig3}, added with random phases $\varphi^s$: $\Omega^s$=[ 2, 4, 8, 16, 32, 64, 128, 256, 512 Hz]$f_{cp}^{-1}$,  amplitudes $\chi^s$=[5, 10, 30, 120, 400, 1000, 1600, 2000, 3000],  and wavevectors  $k_x^s/k_x^0$=[0.2, 0.5, 0.8, 1, 2, 3, 4, 5, 10], where the reference mode $k_x^0$ corresponds to the wave $s=4$ at 16 Hz.
The magnetic field is taken to be 20 nT, corresponding to $f_{cp}=\omega_{cp}/2\pi=0.3$ Hz and $r_c=16\,$km for a 10 eV proton.  The initial $k_x^0 r_c=2$ implies $k_x^0= 1.25\times 10^{-4}\,$m$^{-1}$  ($\lambda=50\,$km). The higher frequency waves $f\sim f_{ce}$ in the  ECD  range are expected to have shorter wavelengths $k_\perp r_e\sim 1$, so we increase $k_x^s$ with frequency to the value $k_x^9/k_x^0=10$ in Equation (\ref{eqh1}). The average spacecraft separation  is about 20 km, so the derived values of $\chi$ for short wavelength ECD waves are likely to be underestimated. We therefore set $\chi^9\approx 3000$.

To transform the numerical results into the satellite frame for a comparison with the measured flux in Figure~\ref{Fig4}, the proton  distribution function is shifted in velocity  by $v_{x0}= 360\,$km\,s$^{-1}$, which corresponds to the proton streaming energy of 0.7 keV. The differential proton energy flux is obtained from the particle distribution by the following procedure: (i) identify the particles within an energy band $K$ to $K+\Delta K$, (ii) add up the differential energy fluxes  $v K/2$ for these particles and divide the sum by $\Delta K$, (iii) divide the result by the total number of particles $M$ and multiply by the background number density $n_p$, and (iv) divide by $4\pi$ to obtain the omnidirectional flux per steradian.

The simulation result in Figure~\ref{Fig5} shows the differential proton energy flux of the initial Maxwellian distribution function (red dashed line), and at times $t=0.5,\,1$ and $1.5\,f_{cp}^{-1}$ (cyclotron periods). After $0.5\,f_{cp}^{-1}$, there is bulk heating and an acceleration of protons up to an energy $\sim 3\,{\rm keV}$ with a low-density tail of protons reaching almost 100 keV (blue solid line), and after this a rapid acceleration of protons up to $\sim 100\,{\rm keV}$ at time $1\,f_{cp}^{-1}$ (black line). At time $1.5 f_{cp}^{-1}$ (magenta) some protons are accelerated above 200 keV.  The simulation demonstrates that the observed proton energy flux in Figure~\ref{Fig4} can be achieved by stochastic acceleration by a wide spectrum of randomly phased waves having increasingly large frequencies, consistent with the concept of the acceleration lane introduced by \citeA{Stasiewicz:2021a}. Similar results (not shown) are also obtained when waves of increasing frequencies are applied in sequence, and not as an ensemble of waves.

The acceleration of protons from 10 eV to 100 keV, i.e., by $10^4$ factor can be achieved in a remarkably short time of one gyroperiod.  The simulation correctly reproduces the power law distribution at lower energies,  however there are some differences at high energies. The presence of humps in the simulated spectrum can be attributed to a discrete number of waves used in the simulations, which have specific phase velocities and would interact resonantly with protons at specific energies. The lowest values of the energy flux distribution at highest energies is determined by 1-count level of test particles, and could be lower if a larger number of particles than $M=10^7$ were used in the simulation. Particles are accelerated in localized bursty/intermittent waves with large $\chi$ values, as indicated by the wavelet decomposition in Figure~\ref{Fig3}.  The heating time of one gyroperiod obtained in the simulation should be regarded as the accumulation of shorter interaction times with bursty waves, so in practice it could be significantly longer.  The simulation is not aimed to provide fitting to the data, but to show qualitatively the feasibility of the discussed mechanism to energize protons to the observed energies of 100 keV.

The large difference in the maximum acceleration of protons between quasi-perpendicular shocks ($K<10\,$keV) and quasi-parallel shocks ($K\sim 100\,$keV) is related to the interaction time with waves \cite{Stasiewicz:2021a}. At quasi-perpendicular shocks, solar wind ions are rapidly convected across the shock and are accelerated only by LHD waves up to the limit (\ref{LHD}), or to the limit (\ref{KE}) computed for lower hybrid waves  ($f<20$ Hz) during a short time comparable to one gyroperiod. This limit is seen also in Figure~\ref{Fig1}(a) after 07:30 UTC for ions transported to the magnetosheath. In quasi-parallel shocks, energetic  ions remain longer time in  regions with  large electric fields by  meandering between the shocklets and bouncing between field compressions in Figure~\ref{Fig2}.  They repetitively interact with higher frequency bursty waves, which    increases their energy to the limit (\ref{KE}) computed for $f<512\,$Hz, as observed in this study during times 06:55-07:25 UTC.

\section{Conclusions}

This investigation shows that protons can be efficiently accelerated up to energies exceeding 100 keV within a few proton cyclotron periods by waves generated by gradient-driven lower hybrid drift instability, followed by the modified two-stream instability and the electron cyclotron drift instability due to the initial electric field generated by the LHD waves. This is supported by recent observations by the MMS spacecraft where large amplitude  drift waves are observed where the conditions for instability is fulfilled, correlated with rapid heating and acceleration of ions. The observations are consistent with the acceleration capacity of LHD waves given by Equation (\ref{LHD}) and the acceleration capacity to the highest energies  of hundreds keV by the ExB mechanism of higher frequency waves in the range $f_{lh}-f_{ce}$ given in Equation (\ref{KE}). Our results imply that the concepts of Fermi acceleration, diffusive shock acceleration, and shock drift acceleration are not needed to explain proton acceleration to hundreds keV observed in MMS data at the Earth's bow shock.

\acknowledgments
B.E. acknowledges support from the EPSRC (UK), grants EP/R004773/1 and EP/M009386/1.
DLT was supported via an International Space Science Institute (ISSI) International Team (Lead: L. B. Wilson III, 2019-2020) and by funding from NASA (Grant NNX16AQ50G and MMS Contract NNG04EB99C at Southwest Research Institute). The data underlying this article are available to the public  through the MMS Science Data Center at
the Laboratory for Atmospheric and Space Physics (LASP),
University of Colorado, Boulder: https://lasp.colorado.edu/mms/sdc/public/. 
We have used the  IRFU-Matlab analysis package available at https://github.com/irfu/irfu-matlab.


\end{document}